\documentclass[12pt]{article}
\usepackage{amsmath, amssymb,graphicx,citesort}

\newwrite\ffile\global\newcount\figno \global\figno=1

\def\writedef#1{}

\input epsf
\def\figin{\epsfcheck\figin}\def\figins{\epsfcheck\figins}
\def\epsfcheck{\ifx\epsfbox\UnDeFiNeD
\message{(NO epsf.tex, FIGURES WILL BE IGNORED)}
\gdef\figin##1{\vskip2in}\gdef\figins##1{\hskip.5in}
\else\message{(FIGURES WILL BE INCLUDED)}%
\gdef\figin##1{##1}\gdef\figins##1{##1}\fi}

\def\figinsert{}
\def\ifig#1#2#3{\xdef#1{fig.~\the\figno}
\writedef{#1\leftbracket fig.\noexpand~\the\figno}%
\figinsert\figin{\centerline{#3}}\medskip\centerline{\vbox{\baselineskip12pt
\advance\hsize by -1truein\center\footnotesize{  Fig.~\the\figno.} #2}}
\bigskip\endinsert\global\advance\figno by1}
\def\endinsert{}

\usepackage{amssymb}
\usepackage{graphics}
\begin{document}
\baselineskip 18pt
\newcommand{\Tr}{\mbox{Tr\,}}
\newcommand{\beq}{\begin{equation}}
\newcommand{\eeq}{\end{equation}}
\newcommand{\bea}{\begin{eqnarray}}
\newcommand{\eea}[1]{\label{#1}\end{eqnarray}}
\renewcommand{\Re}{\mbox{Re}\,}
\renewcommand{\Im}{\mbox{Im}\,}

\def\N{{\cal N}}
\def\one{\hbox{1\kern-.8mm l}}


\thispagestyle{empty}
\renewcommand{\thefootnote}{\fnsymbol{footnote}}

{\hfill \parbox{4cm}{
        SHEP-06-04 \\
}}

\bigskip

\begin{center} \noindent \Large \bf
Gauge Invariant Regularization in the AdS/CFT Correspondence and Ghost
D-branes
\end{center}

\bigskip\bigskip\bigskip

\centerline{ \normalsize \bf Nick Evans, Tim R. Morris and Oliver
J. Rosten \footnote[1]{\noindent \tt
 evans@phys.soton.ac.uk, t.r.morris@soton.ac.uk, o.j.rosten@soton.ac.uk} }

\bigskip
\bigskip\bigskip

\centerline{ \it School of Physics \& Astronomy} \centerline{ \it
Southampton University} \centerline{\it  Southampton, S017 1BJ }
\centerline{ \it United Kingdom}
\bigskip

\bigskip\bigskip

\renewcommand{\thefootnote}{\arabic{footnote}}

\centerline{\bf \small Abstract}
\medskip

{\small \noindent A field theoretic understanding of how the
radial direction in the AdS/CFT Correspondence plays the role of a
{\it gauge invariant} measure of energy scale has long been
missing. In SU(N) Yang-Mills, a realization of a gauge invariant
cutoff has been achieved by embedding the theory in spontaneously
broken SU(N$\mid$N) gauge theory. With the recent discovery of
ghost D-branes an AdS/CFT Correspondence version of this scheme is
now possible. We show that a very simple construction precisely
ties the two pictures together providing a concrete understanding
of the radial RG flow on the field theory side.}

\newpage


The AdS/CFT Correspondence~\cite{Mal,Gub,Wit} is the hugely
successful conjecture of a duality between large N ${\cal N}=4$
Super Yang-Mills theory in four dimensions and type IIB strings /
supergravity on anti-de-Sitter five-space cross a five-sphere. The
field theory is a conformal theory containing an SU(N)$^1$ gauge
field, 6 adjoint scalar  fields, $\phi^i$, and 4 adjoint gauginos,
$\lambda$. Dilatation symmetries in the theory correspond to
rescaling the spatial direction whilst simultaneously rescaling
the fields according to their dimension:

\beq x^\mu \rightarrow \sigma x^\mu, \hspace{1cm} \phi^i \rightarrow
\sigma^{-1} \phi^i, \hspace{1cm}\lambda \rightarrow \sigma^{-3/2}
\lambda.\eeq

This symmetry (part of the full SO(2,4) superconformal symmetry)
matches in the gravity theory to a symmetry of the spacetime
metric

\beq ds^2 = {r^2 \over R^2} dx_4^2 +{R^2 \over r^2} dr^2+ R^2 d
\Omega_5^2, \eeq where $x_4$ are the directions parallel to the D3
world volume and $\Omega_5$ is the metric of a five-sphere. The
radius of the space, $R$, is given by $R^4=4 \pi g_s \mathrm{N}
\alpha'^{2}$ with $g_s$ the string coupling and $\alpha'$ determining
the string tension. We observe that under dilatations in the field
theory the radial direction, $r$, transforms as an energy scale.
This is a crucial part of the standard correspondence with the
radial direction playing the role of renormalization group scale.
However, this identification has always been problematic from the
point of view of the gauge theory. It is well known that it is
very hard to define a gauge invariant energy scale in a field
theory essentially because promoting derivatives to covariant
derivatives makes their Lorentz invariant length depend on the
gauge field.

This problem has been a particularly thorny issue in attempts to
generate a Wilsonian description of renormalization group flow in
gauge theories. If one cannot define a gauge invariant energy
scale how can one follow flow under changes in it? A nice solution
has been proposed in~\cite{a1} and further developed and explored
in~\cite{a2,a3}. In these papers the theory is regularized by
incorporating the theory into an SU(N$\mid$N) gauge theory above
the regularizing scale. This theory has been shown to be so
restrictive that in the large N limit\footnote{In the large N
limit we can effectively ignore the difference between U(N$\mid$N)
and SU(N$\mid$N) and any U(1) factors that otherwise need careful
treatment~\cite{a2}.} there are no interactions above the
regularization scale---the SU(N$\mid$N) theory contains no
dynamics because cancellations between diagrams are exact. The cut
off scale is given by the vev of a scalar field that spontaneously
breaks SU(N$\mid$N) to SU(N)$^2$, thus providing a gauge invariant
regularization for the original SU(N) Yang-Mills. The one-loop and
two-loop $\beta$ function coefficients in Yang-Mills theory have
been reproduced using this regulator and moreover without gauge
fixing~\cite{a2,a3}. These techniques can be extended to general
(perturbative and non-perturbative) calculations in Yang-Mills
theory.

It would be nice to make contact between this approach and the
gauge invariant energy scale of the AdS/CFT Correspondence. We
will be able to make this connection here because
of the recent
discovery of ghost D-branes~\cite{ghost} (also see~\cite{ghost2}):
combining
ghost D-branes
with ordinary D-branes allows the construction of
SU(N$\mid$M) surface gauge theories.
A ghost D-brane is defined by its boundary state being precisely
minus that of an ordinary D-brane---they have negative charges and
tension. Thus, if N ordinary D-branes are coincident with M ghost
D-branes the configuration is the same as if there were N$-$M
ordinary D-branes but with the surface gauge theory being SU(N$\mid$M).

Let us first consider the construction of a 4d
SU(N$\mid$M) ${\cal N}=4$ gauge theory as the surface theory on N D3 branes and
M ghost D3 branes. Each of the ${\cal N}=4$ fields is promoted to a
supergroup field in the adjoint of SU(N$\mid$M). We may quickly
jump to the AdS/CFT Correspondence for this theory by noting that
the supergravity geometry in the near horizon limit around the
stack is just that around N$-$M D3 branes---it is $AdS_5\times S^5$
as written above. The gauge invariant operators must match on to
supergravity fields as in the usual AdS/CFT Correspondence but
this matching is a trivial extrapolation because all the same
fields exist. The operators are the simple extension of the usual
${\cal N}=4$ operators but with the usual trace taken over group
indices replaced by a supertrace taken over
supergroup indices.

The case where M = N is particularly interesting since  then the
geometry is that of no D3 branes! The space is simply flat. This
is clearly the dual of the complete cancellation of dynamics seen
in the field theory for an unbroken SU(N$\mid$N) gauge theory. In
our construction to follow, the appearance of flat space will mark
the onset of a completely regularized theory.

Next it is interesting to consider configurations where the branes
are separated. As usual in the AdS/CFT Correspondence the branes
can be separated in the 6 transverse directions indicating that
the symmetry breaking is associated with vevs for the six scalar
fields, $\phi^i$. Normally these vevs break U(N)$\rightarrow$
U(1)$^N$ as the N D3 branes are separated. Separating the D3s must
still play this role. Similarly separating the M ghost D3 branes
will break U(M)$\rightarrow$ U(1)$^M$. Of more interest though is
the separation of the N D3s and M ghost D3s. If we separate them
as blocks then we are clearly breaking SU(N$\mid$M) $\rightarrow$
SU(N) $\times$ SU(M). This implies the switching on of single
scalar with vev in the supergauge space of the form
\beq \phi =\Lambda \left( \begin{array}{c|c} \one & 0 \\ \hline  0 &
-\one\end{array} \right) + \alpha\Lambda\one,\eeq the
dimensionless $\alpha$ being fixed by the dynamics~\cite{a2}.

The supergravity dual of these set ups will be given by the usual
multi-centre solutions

\beq ds^2 = H^{-1/2} dx_4^2 + H^{1/2} dy^2_6 \eeq but where $H$,
taking into account the ghosts' negative tensions, is given by

\beq H = \sum_{D3} { 4 \pi g_s \alpha' \over \left| y-
y_i\right|^4} - \sum_{ghosts} { 4 \pi g_s \alpha'  \over \left| y-
y_i\right|^4}, \eeq
where $y_i$ are the brane positions.

\begin{figure}[h]
\begin{center}
\includegraphics{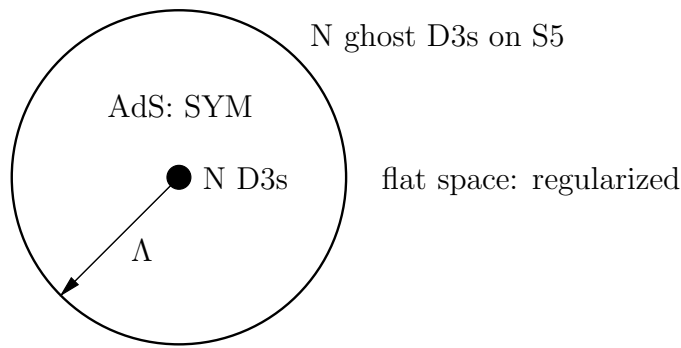}
\caption{Sketch of the brane configuration showing regularized
${\cal N}=4$ SYM. }\label{mVsM}
\end{center}
\end{figure}

We now have all the tools necessary to provide an AdS/CFT
description of using SU(N$\mid$N) to regularize SU(N) ${\cal N}=4$
Yang-Mills. The construction we will use---see figure~\ref{mVsM}---is a stack of N D3 branes
at the origin to generate the field theory we are interested in.
We then surround the D3s at a distance $\Lambda$ by a symmetric
shell distribution of N ghost D3 branes on the surface of a
five-sphere centred on the D3 branes; at large N we may take the ghost
D3 distribution on the five-sphere to be smooth.

The geometry around this set up has two distinct sectors. Within
the five-sphere the space-time is $AdS_5 \times S^5$ because
(essentially by Gauss' Law) the shell does not contribute. Outside
the shell the space is flat since there are the equivalent of no
net D3s contained. The cut off between the two regions is sharp
and discontinuous. The AdS space has been cleanly regularized!

Note that stringy corrections will smooth out this transition.
Indeed, at energies close to $\Lambda$ the effects of the long
strings stretched between the D3 branes and the ghost branes need
to be taken into account. The dynamics of these strings are no
longer negligible.

On the field theory side the low energy theory is just SU(N)
${\cal N}=4$ Yang-Mills, the unphysical ghost SU(N) ${\cal N}=4$
Yang-Mills residing in a decoupled sector (the smearing of the ghost branes 
over the five-sphere breaks the gauge group to U(1)$^\mathrm{N}$), until one moves up to
the scale corresponding to a string of length $\Lambda$. At this
scale the theory becomes the full SU(N$\mid$N) gauge theory and is
regularized. In the field theory this transition is naturally
smooth. Note, however, that covariant higher derivatives are
expected to be required to provide a complete effective
cutoff~\cite{a2}.  Such terms are naturally present in the
effective field theory description of the stringy corrections.

We can change the regularization scale by moving the spherical
shell in the radial direction. It is clear that the radial
position of the sphere precisely corresponds to the symmetry
breaking scale of the supergroup and hence to a gauge invariant
cut off (a precise measure on the field theory side would be the
value of STr $\phi$). This is exactly the identification we
sought to make. It is not clear that this regulator is the only
one that could be used but it does at least provide a clean field
theoretic understanding of the role of the radial distance as a
gauge invariant cut off.

With such a clear gauge invariant regulator in place it should be
possible to make more explicit the link between holographic RG
flow and the Wilsonian exact renormalization group~\cite{kog},
which clearly must be made via its gauge invariant extension~\cite{v2etc,a1}. 
On the field theory side, the AdS/CFT
correspondence is the ideal framework to investigate the
non-perturbative properties of the proposed SU(N$|$N)
regularization~\cite{a1,a2,a3}. It is reasonable to hope that with
an explicit gauge invariant regulator in place, further progress
can be made in understanding non-perturbative aspects of
Yang-Mills itself. Importantly, a description of quarks and QCD is
possible on both sides of the correspondence~\cite{conf,d7s}. Finally, it would be interesting to understand to
what extent the string theory and the low energy quantum gravity
have themselves been regularized in this framework.

\end{document}